\documentclass[twocolumn]{aastex631}
\usepackage{amsmath}

\begin{document}

\title{An eclipsing 8.56 minute orbital period mass-transferring binary
}

\author[0000-0003-4780-4105]{Emma T. Chickles}
\affiliation{Department of Physics, Massachusetts Institute of Technology, Cambridge, MA 02139, USA}
\affiliation{Kavli Institute for Astrophysics and Space Research, Massachusetts Institute of Technology, Cambridge, MA 02139, USA}

\author[0000-0002-0568-6000]{Joheen Chakraborty}
\affiliation{Department of Physics, Massachusetts Institute of Technology, Cambridge, MA 02139, USA}
\affiliation{Kavli Institute for Astrophysics and Space Research, Massachusetts Institute of Technology, Cambridge, MA 02139, USA}

\author[0000-0002-7226-836X]{Kevin B. Burdge}
\affiliation{Department of Physics, Massachusetts Institute of Technology, Cambridge, MA 02139, USA}
\affiliation{Kavli Institute for Astrophysics and Space Research, Massachusetts Institute of Technology, Cambridge, MA 02139, USA}

\author[0000-0003-4236-9642]{Vik S. Dhillon}
\affiliation{Astrophysics Research Cluster, School of Mathematical and Physical Sciences, University of Sheffield, Sheffield S3 7RH, UK}
\affiliation{Instituto de Astrof\'{i}sica de Canarias, E-38205 La Laguna, Tenerife, Spain} 

\author[0000-0002-2218-2306]{Paul Draghis}
\affiliation{Department of Physics, Massachusetts Institute of Technology, Cambridge, MA 02139, USA}
\affiliation{Kavli Institute for Astrophysics and Space Research, Massachusetts Institute of Technology, Cambridge, MA 02139, USA}

\author[0000-0002-6871-1752]{Kareem El-Badry}
\affiliation{Division of Physics, Mathematics and Astronomy, California Institute of Technology, Pasadena, CA, USA}

\author[0000-0002-0948-4801]{Matthew J. Green}
\affiliation{Max-Planck-Institut f\"{u}r Astronomie, K\"{o}nigstuhl 17, D-69117 Heidelberg, Germany}
\affiliation{Homer L. Dodge Department of Physics and Astronomy, University of Oklahoma, 440 W. Brooks Street, Norman, OK 73019, USA}
\affiliation{JILA, University of Colorado and National Institute of Standards and Technology, 440 UCB, Boulder, CO 80309-0440, USA}

\author[0000-0002-5812-3236]{Aaron Householder}
\affiliation{Department of Earth, Atmospheric and Planetary Sciences, Massachusetts Institute of Technology, Cambridge, MA 02139, USA}
\affiliation{Kavli Institute for Astrophysics and Space Research, Massachusetts Institute of Technology, Cambridge, MA 02139, USA}

\author[0000-0002-7332-2751]{Sarah Hughes}
\affiliation{Department of Physics, Massachusetts Institute of Technology, Cambridge, MA 02139, USA}
\affiliation{Kavli Institute for Astrophysics and Space Research, Massachusetts Institute of Technology, Cambridge, MA 02139, USA}

\author[0000-0002-7191-4403]{Christopher Layden}
\affiliation{Department of Physics, Massachusetts Institute of Technology, Cambridge, MA 02139, USA}
\affiliation{Kavli Institute for Astrophysics and Space Research, Massachusetts Institute of Technology, Cambridge, MA 02139, USA}

\author[0000-0001-7221-855X]{Stuart P. Littlefair}
\affiliation{Astrophysics Research Cluster, School of Mathematical and Physical Sciences, University of Sheffield, Sheffield S3 7RH, UK}

\author{James Munday}
\affiliation{Department of Physics, University of Warwick, Coventry CV4 7AL, UK}

\author[0000-0003-4615-6556]{Ingrid Pelisoli}
\affiliation{Department of Physics, University of Warwick, Coventry CV4 7AL, UK}

\author[0009-0005-8281-1644]{Maya S. Redden}
\affiliation{Department of Physics, Stanford University, Stanford, CA 94305, USA}

\author[0000-0003-2858-9657]{John Tonry}
\affiliation{Institute for Astronomy, University of Hawaii, 2680 Woodlawn Drive, Honolulu, HI 96822-1897, USA}

\author[0000-0002-2626-2872]{Jan van Roestel}
\affiliation{Institute of Science and Technology Austria, Am Campus 1, 3400 Klosterneuburg, Austria}
\affiliation{Anton Pannekoek Institute for Astronomy, University of Amsterdam, 1090 GE Amsterdam, The Netherlands}

\author[0009-0008-8658-2764]{F. Elio Angile}
\affiliation{Kavli Institute for Astrophysics and Space Research, Massachusetts Institute of Technology, Cambridge, MA 02139, USA}

\author[0000-0002-3316-7240]{Alex J. Brown}
\affiliation{Hamburger Sternwarte, University of Hamburg, Gojenbergsweg 112, 21029 Hamburg, Germany}

\author[0000-0002-5870-0443]{Noel Castro Segura}
\affiliation{Department of Physics, University of Warwick, Coventry CV4 7AL, UK}

\author[0000-0002-6401-778X]{Jack Dinsmore}
\affiliation{Department of Physics, Stanford University, Stanford, CA 94305, USA}

\author[0000-0003-3665-5482]{Martin Dyer}
\affiliation{Astrophysics Research Cluster, School of Mathematical and Physical Sciences, University of Sheffield, Sheffield S3 7RH, UK}
\affiliation{Research Software Engineering, University of Sheffield, Sheffield S1 4DP, UK}

\author[0000-0001-8467-9767]{Gabor Furesz}
\affiliation{Kavli Institute for Astrophysics and Space Research, Massachusetts Institute of Technology, Cambridge, MA 02139, USA}

\author{Michelle Gabutti}
\affiliation{Kavli Institute for Astrophysics and Space Research, Massachusetts Institute of Technology, Cambridge, MA 02139, USA}

\author[0009-0007-5535-3312]{James Garbutt}
\affiliation{Astrophysics Research Cluster, School of Mathematical and Physical Sciences, University of Sheffield, Sheffield S3 7RH, UK}

\author[0000-0003-1361-985X]{Juliana García-Mejía}
\altaffiliation{51 Pegasi B Fellow, MIT Pappalardo Physics Fellow}
\affiliation{MIT Kavli Institute for Astrophysics and Space Research, Massachusetts Institute of Technology, 77 Massachusetts Ave, Cambridge, MA 02139, USA}
\affiliation{MIT Department of Physics, 77 Massachusetts Ave., Cambridge, MA 02139, USA}
\affiliation{Center for Astrophysics \textbar\ Harvard \& Smithsonian, 60 Garden Street, Cambridge, MA 02138, USA}

\author[0009-0004-3067-2227]{Daniel Jarvis}
\affiliation{Astrophysics Research Cluster, School of Mathematical and Physical Sciences, University of Sheffield, Sheffield S3 7RH, UK}

\author[0000-0001-6894-6044]{Mark R. Kennedy}
\affiliation{School of Physics, University College Cork, Cork, T12 K8AF, Ireland}

\author{Paul Kerry}
\affiliation{Astrophysics Research Cluster, School of Mathematical and Physical Sciences, University of Sheffield, Sheffield S3 7RH, UK}

\author[0000-0003-1631-4170]{James McCormac}
\affiliation{Department of Physics, University of Warwick, Coventry CV4 7AL, UK}

\author[0000-0001-6331-112X]{Geoffrey Mo}
\affiliation{Division of Physics, Mathematics and Astronomy, California Institute of Technology, Pasadena, CA 91125, USA}
\affiliation{Carnegie Observatories, 813 Santa Barbara Street, Pasadena, CA 91101, USA}

\author[0000-0003-0412-9664]{Dave Osip}
\affiliation{Las Campanas Observatory, Carnegie Institution for Science, Colina el Pino, Casilla 601 La Serena, Chile}

\author[0000-0002-2695-2654]{Steven Parsons}
\affiliation{Astrophysics Research Cluster, School of Mathematical and Physical Sciences, University of Sheffield, Sheffield S3 7RH, UK}

\author{Eleanor Pike}
\affiliation{Astrophysics Research Cluster, School of Mathematical and Physical Sciences, University of Sheffield, Sheffield S3 7RH, UK}

\author[0000-0002-9602-2217]{Jack Piotrowski}
\affiliation{Carnegie Observatories, 813 Santa Barbara Street, Pasadena, CA 91101, USA}

\author[0000-0001-6711-3286]{Roger W. Romani}
\affiliation{Department of Physics, Stanford University, Stanford, CA 94305, USA}

\author[0000-0002-0403-1547]{David Sahman}
\affiliation{Astrophysics Research Cluster, School of Mathematical and Physical Sciences, University of Sheffield, Sheffield S3 7RH, UK}


\author{Rob Simcoe}
\affiliation{Department of Physics, Massachusetts Institute of Technology, Cambridge, MA 02139, USA}
\affiliation{Kavli Institute for Astrophysics and Space Research, Massachusetts Institute of Technology, Cambridge, MA 02139, USA}




\begin{abstract}
We report the discovery of ATLAS J101342.5$-$451656.8 (hereafter referred to as ATLAS~J1013$-$4516), an 8.56 minute orbital-period mass-transferring AM~Canum~Venaticorum (AM~CVn) binary with a mean \textit{Gaia} magnitude of $G=19.51$, identified via periodic variability in light curves from the Asteroid Terrestrial-impact Last Alert System (ATLAS) of \textit{Gaia} white-dwarf candidates. Follow-up with the Large Lenslet Array Magellan Spectrograph (LLAMAS) shows a helium-dominated accretion disk, and high-speed ULTRACAM photometry reveals pronounced primary and secondary eclipses. We construct a decade-long timing baseline leveraging lightcurves from the ATLAS and \textit{Gaia} surveys, as well as the high-speed imagers ULTRACAM on the NTT and proto-Lightspeed on the Magellan Clay telescope. From this timing baseline, we measure an orbital period derivative of $\dot{P}=-1.60\pm0.07\times 10^{-12}$ s s$^{-1}$. Interpreted in the context of stable mass transfer, the magnitude and sign of $\dot{P}$ indicate that the orbital evolution is governed by the interplay between gravitational-wave--driven angular-momentum losses and mass transfer, directly probing the donor's structural response to mass loss. We constrain the accretor and donor mass based on stable mass-transfer arguments assuming angular momentum loss dominated by gravitational wave emission, allowing us to infer the characteristic gravitational wave strain of the binary for future space-based GW observatories such as the Laser Interferometer Space Antenna (\textit{LISA}). We predict a characteristic strain corresponding to a 4-year \textit{LISA} signal-to-noise ratio $\gtrsim10$, establishing ATLAS~J1013$-$4516 as a strong prospective \textit{LISA} source that will probe long-term orbital evolution in the mass-transferring regime.
\end{abstract}

\keywords{White dwarf stars(1799) --- Compact binary stars(297) --- Gravitational wave sources(677) --- Time domain astronomy(2109)}


\section{Introduction} \label{sec:intro}

AM~Canum~Venaticorum (AM~CVn) binaries are a rare class of ultracompact binary systems in which a white dwarf accretes helium-rich material from a degenerate or semi-degenerate donor. Their orbital periods range from about 5 to 70 minutes \citep{2018A&A...620A.141R}, and their blue continua, strong He I/He II emission, and absence of Balmer lines distinguish them from hydrogen-accreting cataclysmic variables (see \citealt{2010PASP..122.1133S} for a comprehensive review and \citealt{2025A&A...700A.107G} for a recent catalog).  

AM~CVns are valuable laboratories for studying the stability of mass transfer, angular-momentum loss through gravitational radiation, and the physics of binary evolution at ultra-short orbital periods. They are also persistent millihertz gravitational-wave sources detectable by the upcoming Laser Interferometer Space Antenna (LISA; \citealt{2017arXiv170200786A}), and measurements of their orbital-period derivatives directly probe the balance between angular momentum losses from gravitational-wave emission and orbital expansion driven by mass-transfer. Such measurements have been achieved for only a few ultracompact binaries with significant, non-zero orbital-period derivatives, including a small subset of disk-accreting systems such as ES~Ceti ($P=10.3$~min; \citealt{2018ApJ...852...19D}) and two recently discovered systems with $P=7.95$ and $8.68$~min presented in \citet{2024ApJ...977..262C}.

To date, only $\sim$100 AM~CVn binaries have been identified, and the majority have orbital periods $\gtrsim20$~min \citep{2010PASP..122.1133S,2025A&A...700A.107G}. The known sample has been assembled primarily through a combination of spectroscopy (identifying blue continua with strong He~I/He~II features and no Balmer lines), the detection of dwarf nova outbursts in transient surveys, and searches for periodic variability in time-domain data, and X-ray--selected ultracompact binaries \citep[e.g.,][]{1999A&A...349L...1I,2010PASP..122.1133S,2020ApJ...905...32B,2025A&A...700A.107G}. The handful of AM~CVn binaries with periods below $\sim10$ min lie where gravitational-wave--driven angular-momentum loss is strongest, making them some of the highest--strain-amplitude Galactic sources in the millihertz band. Many are expected to be individually resolved by LISA, and their numbers and period distribution inform models of the unresolved Galactic gravitational-wave foreground \citep{2017MNRAS.470.1894K,2018MNRAS.480..302K}. These ultracompact, mass-transferring binaries also probe helium mass-transfer rates and stability in double-degenerate binaries, providing empirical constraints on the conditions that separate stable accretion from the regimes that produce helium-powered thermonuclear transients \citep{Webbink84,2007ApJ...662L..95B,2009ApJ...699.1365S,2014MNRAS.445.3239P,Munday25}. The advent of wide-field, high-cadence surveys---including the Asteroid Terrestrial-impact Last Alert System (ATLAS; \citet{2018PASP..130f4505T}), the Zwicky Transient Facility (ZTF; \citet{2019PASP..131a8002B}), the \emph{Gaia} mission \citep{2016A&A...595A...1G}, and the Transiting Exoplanet Survey Satellite (\emph{TESS}; \citet{2015JATIS...1a4003R})---now enables systematic searches for such short-period sources across the sky \citep{2020ApJ...905...32B, 2025A&A...700A.107G}.

We conducted a variability search of $\sim$1.3~million \emph{Gaia} white-dwarf candidates \citep{2021MNRAS.508.3877G} using ATLAS light curves, leading to the discovery of ATLAS~J101342.5$-$451656.8 (hereafter ATLAS~J1013$-$4516), an 8.56~min AM~CVn binary exhibiting deep eclipses and a measurable orbital-period derivative. In this paper, we present its discovery, multiwavelength characterization, and implications for mass transfer and gravitational-wave evolution. Section \ref{sec:obs} describes the observations, Section \ref{sec:analysis} presents the spectral, photometric, and timing analysis, Section \ref{sec:discussion} discusses the system's physical interpretation, orbital evolution, and population context, and Section \ref{sec:conclusion} summarizes our conclusions.

\section{Observations} \label{sec:obs}

We combine survey photometry from ATLAS and \emph{Gaia}, follow-up spectroscopy, and high-speed optical and X-ray observations to characterize ATLAS~J1013$-$4516 across multiple wavelengths and timescales. Below we describe the ATLAS discovery light curves, \emph{Gaia} epoch photometry, spectroscopic follow-up with Magellan/LLAMAS, high-speed photometry with ULTRACAM and \textit{proto-Lightspeed}, and constraints from archival X-ray observations with \textit{XMM-Newton}.

\subsection{ATLAS Photometry and Discovery}
\label{sec:atlas}

As part of a systematic search for short-period variability among \emph{Gaia}-selected white dwarfs \citep{2021MNRAS.508.3877G}, we analyzed $\sim$1.3 million ATLAS light curves from the Asteroid Terrestrial-impact Last Alert System \citep{2018PASP..130f4505T}, spanning observations through early 2022. Periods were identified using the Box-Least-Squares (BLS) algorithm \citep{2016ascl.soft07008K}, optimized for detecting eclipsing signals, with an oversampling factor of 3. The ATLAS light curves of ATLAS~J1013$-$4516 exhibit a highly significant BLS peak at a period of $\sim$8.56~min, rising $\sim$80 median absolute deviations above the median periodogram level. The amplitude and short timescale of the modulation prompted spectroscopic and high-speed photometric follow-up, described in subsequent sections.

For visualization and timing analysis, we additionally retrieved updated ATLAS forced-photometry light curves for ATLAS~J1013$-$4516 extending to the most recent available epochs. These light curves were obtained using PSF photometry performed directly on the survey images. We used the \textit{reduced} photometry mode, which avoids discontinuities that can arise in difference-image light curves when reference templates are updated. The combined ATLAS dataset spans observations from 2016 onward and includes 461 $c$-band and 1601 $o$-band measurements, typically sampled every $\sim$2--4 nights, with median uncertainties of 0.27~mag and 0.73~mag, respectively. After removing points fainter than the five-sigma limiting magnitude and applying sigma clipping independently to each filter, 427 ($c$) and 1511 ($o$) observations remain.

\begin{figure}[ht!]
    \includegraphics[width=0.45\textwidth]{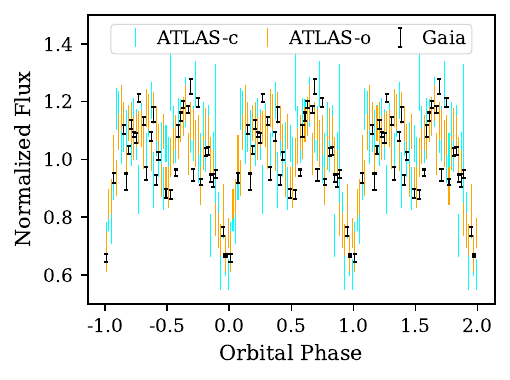}
    \caption{Phase-folded light curves of ATLAS J1013$-$4516 from synoptic surveys, folded on the timing ephemeris listed in Table~\ref{tab:inferred} with phase zero defined at mid-eclipse. Cyan ($c$, 420–650\,nm) and orange ($o$, 560–820\,nm) points show phase-binned ATLAS photometry, overlaid with \emph{Gaia} epoch photometry. All data are binned using inverse-variance--weighted averaging and repeated over three orbital cycles for visualization. The ATLAS light curves are linearly rescaled to match the interquartile range and median level of the phase-binned \emph{Gaia} photometry. The resulting $\sim$50\% eclipse-like attenuation and coherent periodic modulation motivated high-speed follow-up.}
    \label{fig:atlas}
\end{figure}

When phase-folded on the orbital period, the ATLAS light curves (Figure~\ref{fig:atlas}) show a coherent, non-sinusoidal waveform with a narrow, eclipse-like dip at a consistent phase (and a possible secondary minimum), suggestive of binarity. Although the median flux in both bands is slightly negative---consistent with a small photometric zero-point offset in the ATLAS forced-photometry pipeline for this crowded Galactic-plane field (1.86\arcsec\ pixels; $\sim$2-pixel FWHM)---this offset does not affect relative variability measurements or the timing coherence of the detected signal. However, the ATLAS data alone do not uniquely determine the physical origin of the variability; alternative scenarios such as hotspot-dominated emission in magnetic systems cannot be excluded without spectroscopy and higher-cadence, multi-band photometric observations.

\subsection{Gaia Epoch Photometry}

In addition to ATLAS, we make use of time-resolved photometry from the \emph{Gaia} mission \citep{2016A&A...595A...1G}. We retrieve \emph{Gaia} G-band epoch photometry from Data Release~3 \citep{2023A&A...674A...1G}, which provides 54 sparsely sampled but high-precision flux measurements spanning $\sim$2014--2017. The astrometric parallax of ATLAS~J1013$-$4516 is poorly constrained ($0.1992\pm0.2808$~mas) and is not used to infer the system distance.

When phase-folded on the orbital period, the \emph{Gaia} photometry aligns coherently with the modulation observed in the ATLAS light curves (Figure~\ref{fig:atlas}), providing an independent confirmation of the periodic signal and contributing additional epochs to the long-baseline timing analysis (Section~\ref{sec:oc}).





\subsection{LLAMAS}

We obtained phase-resolved spectroscopy of ATLAS~J1013$-$4516 using the Large Lenslet Array Magellan Spectrograph (LLAMAS) on the Magellan Baade Telescope at Las Campanas Observatory on 2025 December 14. We acquired 28 consecutive 30\,s 
exposures covering 3,400-10,000 \AA\ at a resolving power $R\approx2000$. Given the 8.56\,min orbital period, this sequence samples the full orbit, with each exposure corresponding to $\approx 1/17$ of an orbital cycle. Conditions were clear with seeing of 0.5--0.7\arcsec. 

We reduced the LLAMAS observations using the in-development \texttt{llamas-pyjamas} pipeline\footnote{\url{https://github.com/mit-kavli-institute/llamas-pyjamas/tree/main/llamas_pyjamas}}, an end-to-end framework that currently outputs 2D row-stacked spectra (RSS) and preliminary 3D data cubes \citep{Hughes2026}. Because modules for per-fiber throughput normalization, sky subtraction, and absolute flux calibration remain under active development, we supplemented the automated reduction with a small number of manual corrections to address these steps.

\begin{figure*}[ht!]
    \center \includegraphics[width=\textwidth]{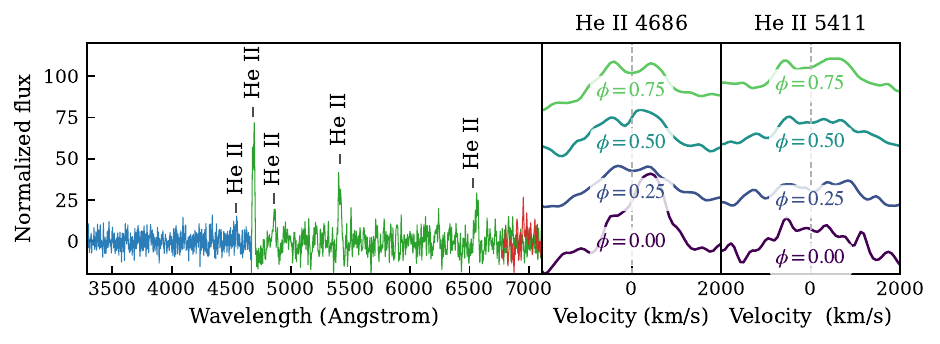}
    \caption{Coadded, continuum-subtracted Magellan/LLAMAS spectra of ATLAS~J1013$-$4516 obtained on 2025 December 14. The displayed wavelength range is restricted to the highest signal-to-noise, non--sky-dominated region of the LLAMAS coverage. Right panels show phase-resolved He\,\textsc{ii} line profiles, constructed by binning the spectra into four equal orbital phase bins and arbitrarily offset vertically for clarity. Both the phase-binned line profiles and the coadded spectrum have been smoothed with a Gaussian kernel for visualization.}
    \label{fig:LLAMAS}
\end{figure*}

The resulting coadded spectrum (Figure~\ref{fig:LLAMAS}) exhibits strong He II emission characteristic of an AM~CVn accretion disk. The spectra are shown in the observatory frame; no barycentric or systematic velocity correction has been applied. We discuss the LLAMAS spectral properties in more detail in Section~\ref{sec:spectra}.

\subsection{ULTRACAM}

We obtained high-speed photometry of ATLAS J1013$-$4516 using ULTRACAM \citep{2007MNRAS.378..825D}, a triple-beam CCD camera mounted on the 3.5\,m New Technology Telescope (NTT) at La Silla Observatory. Observations were carried out on 2023 March 29-30, 2024 February 6-7, 2024 December 30, 2025 February 20, and 2025 March 28, with simultaneous imaging in the Sloan $u_S$, $g_s$, and red ($r_s$ or $i_s$) filters.

\begin{figure*}[ht!]
    \centering
    \includegraphics[width=\textwidth]{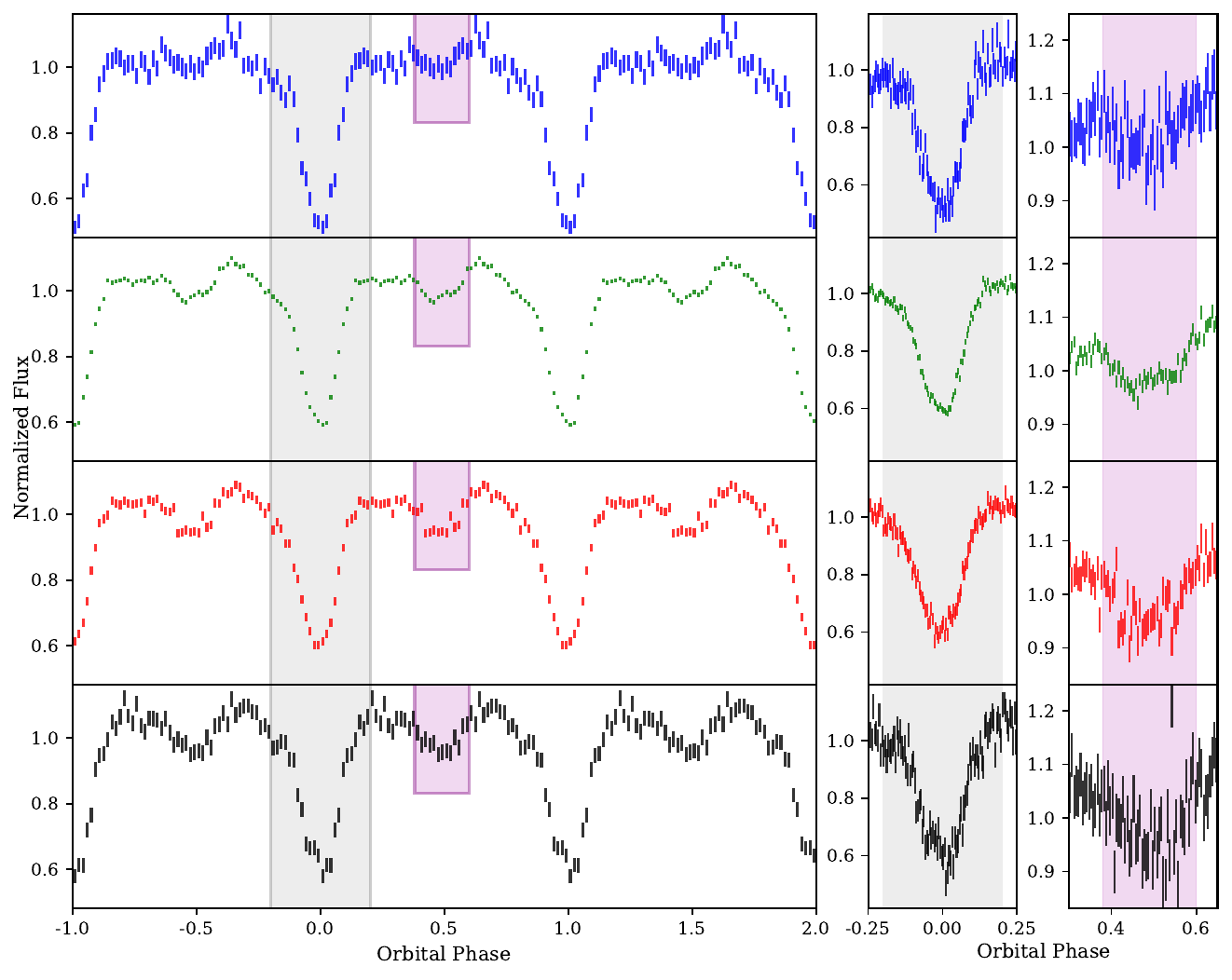}
    \caption{ULTRACAM light curves of ATLAS J1013$-$4516, folded on the orbital period. The $u_s$, $g_s$, $r_s$, and $i_s$ light curves shown here are constructed from multiple nights of observations using different filter triplets ($u_sg_sr_s$ and $u_sg_si_s$ configurations). Central wavelengths are 3520, 4729, 6196, and 7708\,\AA for $u_s$, $g_s$, $r_s$, $i_s$, respectively. 
    Each panel includes insets (right columns) that zooms in on the primary and secondary eclipses. Gray and purple shaded regions mark the primary and secondary eclipse windows, respectively.
    The insets highlight the wavelength dependence of the eclipse depths: the primary eclipse depends towards shorter wavelengths, reflecting the dominance of a hot, blue accretion component, while the secondary eclipse becomes more pronounced at longer wavelengths as the cooler, red-emitting donor contributes more strongly. Fluxes are normalized by the out-of-eclipse median.}
    \label{fig:ultracam}
\end{figure*}

The data were reduced using the \texttt{hipercam} pipeline\footnote{\url{https://github.com/HiPERCAM/hipercam}}. The reduction included bias subtraction, flat-field correction, and aperture photometry. Differential light curves were obtained by dividing the target flux by that of a nearby comparison star. Several comparison stars were tested to confirm that the results were not affected by variability in the reference source. The phase-folded ULTRACAM light curves (Figure \ref{fig:ultracam}) reveal deep primary and shallower secondary eclipses across all bands. These observations confirm the 8.56-min orbital period and provide multi-band constraints on the system's temperature structure.

\subsection{Proto-Lightspeed}
To further resolve short-timescale structure in the eclipse and to extend the eclipse-timing baseline, we obtained high-speed photometric observations of ATLAS J1013$-$4516 using the proto-\textit{Lightspeed} camera on the Magellan Clay Telescope at Las Campanas Observatory. Proto-\textit{Lightspeed} is a high-speed optical imager currently under commissioning for the Magellan telescopes \citep{Layden2026}. The instrument employs a Hamamatsu ORCA-Quest 2 CMOS image sensor with deep sub-electron readout noise, allowing high-cadence photometry of this 19th-magnitude system without significant readout-noise penalties. During our September 2025 observations, we obtained 218 consecutive 5-s exposures in the Sloan $g^{\prime}$ band over a 20-min sequence. During our December 2025 observations, we obtained an additional $\sim10,000$ consecutive 1-s exposures in the same band over a 3-hour sequence.

The instrument timing system is synchronized to a GPS reference. For the September commissioning data, the clock was initialized against GPS at the start of each night but was not continuously locked; commissioning observations of the Crab Pulsar demonstrate that the absolute timing remained stable to a fraction of a millisecond over several hours, well below the timing uncertainties associated with measuring the orbital phase of the light curve. Frame-to-frame (relative) timing is stable at the sub-microsecond level. For the December observations, continuous GPS locking was enabled, providing absolute timing accuracy better than 10 microseconds throughout the observing sequence. Data were reduced with a custom aperture-photometry pipeline.

These observations yield a precise mid-eclipse time for ATLAS J1013$-$4516, which we incorporate into the global timing solution to refine the system's ephemeris (see Section \ref{sec:oc}).

\subsection{XMM-Newton}
An archival pointed \textit{XMM-Newton} observation obtained on 2002 May 30 (ObsID~0112880101) serendipitously covered the field of ATLAS~J1013$-$4516, yielding a deep non-detection in X-rays. We estimated flux upper limits using the \textit{XMM-Newton} Upper Limit Server (ULS; \citealt{2020A&A...641A.136W}), which models the local background and instrument response to compute $3\sigma$ limits at a specified sky position. For an effective exposure time of 6.2~ks, we obtain $3\sigma$ flux upper limits of $F_{0.2-2\,\mathrm{keV}} < 8.4\times10^{-15}\,\mathrm{erg\,s^{-1}\,cm^{-2}}$, $F_{2-12\,\mathrm{keV}} < 5.1\times10^{-14}\,\mathrm{erg\,s^{-1}\,cm^{-2}}$, and $F_{0.2-12\,\mathrm{keV}} < 2.9\times10^{-14}\,\mathrm{erg\,s^{-1} cm^{-2}}$, assuming a power-law spectral model with photon index $\Gamma = 1.5$ and Galactic absorption of $N_{\mathrm{H}} = 1\times10^{21}\,\mathrm{cm^{-2}}$. The choice of $\Gamma=1.5$ reflects the discrete spectral options available in the ULS interface and is broadly consistent with the hard X-ray spectrum ($\Gamma\simeq1.3$) measured for the high-inclination, disk-accreting 13.7-min white dwarf binary ZTF~J1901$+$5309 using \textit{Chandra} \citep{2023ApJ...953L...1B}.

\section{Analysis}
\label{sec:analysis}

\subsection{Optical Spectral Features} \label{sec:spectra}

The coadded Magellan/LLAMAS spectrum of ATLAS~J1013$-$4516 (Figure~\ref{fig:LLAMAS}) is dominated by strong He\,\textsc{ii} emission, most prominently at 4542, 4686, 5411, and 6560\,\AA, the latter identified as the He\,\textsc{ii} Pickering transition rather than H$\alpha$ based on the simultaneous presence of the He\,\textsc{ii} 5411\,\AA line at comparable strength. The phase-resolved line profiles (Figure~\ref{fig:LLAMAS}, right panels) exhibit persistent double-peaked morphologies consistent with Doppler broadening in a rotating accretion disk. Mild asymmetries in the relative peak strengths are visible at some orbital phases (e.g., $\phi=0.0$ and $\phi=0.5$ in the He\,\textsc{ii}~4686~\AA profile), indicating departures from a perfectly axisymmetric brightness distribution. However, within the signal-to-noise and spectral resolution of the LLAMAS data, we do not detect a coherent phase-dependent velocity modulation or a distinctive S-wave component. This implies that any localized emission regions (e.g., a stream--disk impact or azimuthal disk structure) are weak, spatially extended, or diluted by the dominant disk emission. No Balmer emission or absorption features are detected, indicating a helium-dominated accretion flow. 

Several weak features are present near the expected wavelengths of He\,\textsc{i} transitions (e.g., 4471, 5015, 5876, and 6678~\AA), but their low amplitudes and limited signal-to-noise preclude a robust identification. The relative prominence of He\,\textsc{ii} over He\,\textsc{i} suggests a hot, high-excitation environment, as seen in the most luminous AM~CVn systems such as ES~Cet \citep{2005PASP..117..189E} and ZTF~J0546+3843 \citep{2024ApJ...977..262C}. 

Weak emission near $\sim$3995\,\AA\ and the 4630--4640\,\AA\ region may correspond to N\,\textsc{ii} transitions, though these features are detected at marginal significance. No statistically significant C or O emission lines are observed. Higher signal-to-noise spectroscopy will be required to place meaningful constraints on the elemental abundances. We now turn to the time-domain data, which constrain the system's orbital geometry and reveal non-axisymmetric emission components that are not cleanly separable in the phase-averaged spectrum.

\subsection{Light Curve Morphology and Modeling} \label{sec:lcurvegeometry}

\begin{figure}[ht!]
    \raggedright\includegraphics[width=0.48\textwidth]{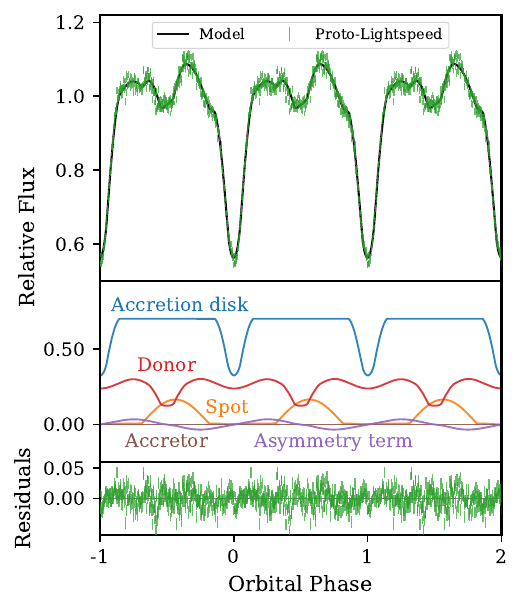}
    \caption{Phase-folded proto-Lightspeed \emph{g'}-band light curve of the AM CVn system ATLAS~J1013$-$4516 (top; green points), repeated over three orbital cycles for visualization. The top panel shows binned photometry (green) and the best-fitting \texttt{lcurve} model overplotted (black). The bottom panel shows the corresponding additive model components, including the accretion disk (blue), donor star (red), accretor (brown), hot-spot contributions (orange), and an asymmetry term (purple). The sum of these components reproduces the total model light curve shown in the top panel.
    \label{fig:lcurve}}
\end{figure}

To provide an illustrative interpretation of the variability of ATLAS~J1013$-$4516, we fit simple \texttt{lcurve} models to the phase-folded and binned \emph{g'}-band proto-Lightspeed light curve using a minimal geometric prescription (Figure~\ref{fig:lcurve}; \citealt{2010MNRAS.402.1824C}). The proto-Lightspeed data offer superior time resolution during eclipse, making them particularly well suited for demonstrating the relative contributions of donor and accretion-related structures to the observed light curve.

The model includes eclipses of the accretion disk and donor, ellipsoidal modulation of the donor, and a hot spot associated with the stream-disk impact. However, these standard components alone were insufficient to reproduce a subtle but coherent out-of-eclipse asymmetry present in the data. To capture this feature, we therefore include an additional phenomenological sinusoidal term. This component is intended to account for departures from axisymmetry that are not explicitly represented in the simplified disk and hot-spot geometry.

Model parameters were guided toward the observed morphology using standard optimization routines available within \texttt{lcurve}, including Levenberg--Marquardt and downhill simplex algorithms. This fitting procedure was employed iteratively to obtain a numerically well-matched representation of the light curve, rather than to derive a unique or fully self-consistent physical solution. In particular, the inclusion of the phenomenological asymmetry term precludes a strictly physical interpretation of the fitted parameters. A fully physical fit would require replacing this descriptive component with a self-consistent model for azimuthal disk structure, which is beyond the scope of this work.

Despite the non-uniqueness of the model, it reproduces the principal morphological features of the data: a deep primary eclipse driven by the obscuration of the accretion disk, the absence of detectable accretor flux, and a modest blue-band hump outside eclipse. Together, these features indicate a high-inclination system in which the optical light is dominated by disk emission and accretion-related structures, consistent with expectation for a short-period AM~CVn binary.

In addition to the morphological modeling, we directly measured the wavelength dependence of the secondary-eclipse depth using ULTRACAM multi-band photometry. For each filter ($u_S$, $g_S$, $r_S$, and $i_S$), we computed weighted mean fluxes in in-eclipse and out-of-eclipse phase intervals, obtaining fractional eclipse depths $\delta_i\equiv\frac{F_{\rm out}-F_{\rm in}}{F_{\rm out}}$ of $1.6\pm1.2\%, 3.3\pm0.3\%, 6.2\pm0.6\%, 7.4\pm1.0\%$, respectively. The eclipse depth increases monotonically toward longer wavelengths, indicating that the eclipsed component is redder compared to the dominant out-of-eclipse light.

Importantly, the measured eclipse depths do not directly correspond to the fractional flux contribution of the donor star, as they are degenerate with the occultation fraction $f_{\rm occ}$ of the donor that is obscured during eclipse. The light-curve modeling favors a donor that is vertically extended relative to the thin accretion disk, implying $f_{\rm occ}<1$. We use these eclipse-depth measurements in the spectral energy distribution analysis (\ref{sec:sed}), where they are incorporated into a joint likelihood alongside the broadband photometry.

\begin{figure*}[ht!]
    \center
    \includegraphics[width=0.99\textwidth]{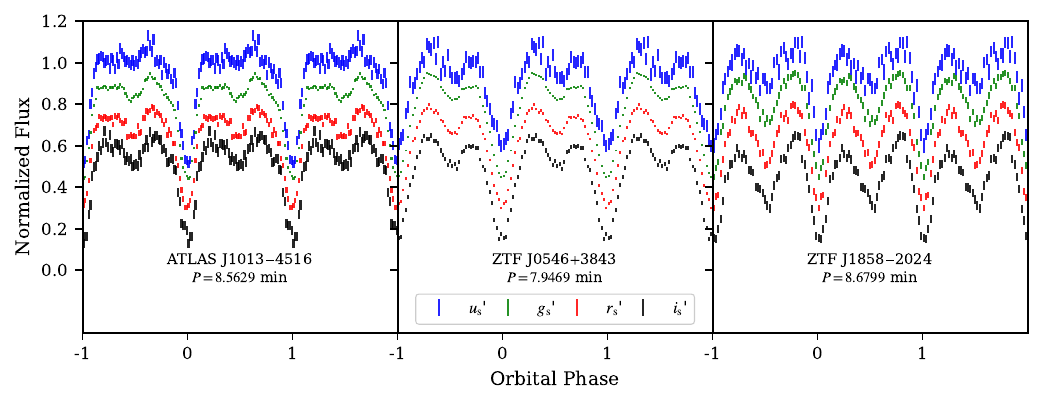} 
    \caption{Comparison of ULTRACAM light curves of ATLAS J1013$-$4516 with short-period, disk-accreting AM\,CVn binaries ZTF J0546+3843 and ZTF J1858+2024. Each row shows simultaneous multi-band photometry phase-folded on the orbital period. The dashed line marks the phase of maximum $u$ flux. All three systems display deep disk eclipses, but the phase and amplitude of the blue hump differ, reflecting variations in hot-spot geometry and stream-impact location within the disk.}
    \label{fig:amcvnlc}
\end{figure*}

The light curve morphology of ATLAS~J1013$-$4516 can be placed in context by comparison with other short-period, disk-accreting AM~CVn binaries (Figure~\ref{fig:amcvnlc}). Among systems in this regime, systematic differences are observed in the orbital phase of the blue-band hump. In ATLAS~J1013$-$4516 and ZTF~J1858$+$2024, the hump reaches maximum brightness after the secondary eclipse, whereas in ZTF~J0546$+$3843 it occurs prior to secondary eclipse \citep{2024ApJ...977..262C}. These phase offsets indicate differences in the azimuthal location of enhanced emission within the accretion disk. 

A natural explanation for these differences is variation in disk size relative to the ballistic stream trajectory. In systems with more compact disks, the stream can travel farther along its ballistic path before encountering the disk, allowing greater Coriolis deflection and producing a hot-spot impact on the far side of the disk that appears at later orbital phase. Conversely a larger disk would intercept the stream closer to the donor's Lagrange point, yielding a hot spot at earlier orbital phase. Such geometric differences can arise even among systems with similar orbital periods and do not require changes in the ballistic stream trajectory itself. 

The post-eclipse hump in ATLAS~J1013$-$4516 may also signal contributions from a vertically extended multi-site emission associated with the stream--disk interaction. In hydrogen-accreting cataclysmic variables, partial stream overflow---where material skims the outer disk rim and later re-impacts the disk at smaller radii---can produce secondary emission regions displaced in azimuth from the primary hot spot \citep{1999ApJ...510..867H, 2002ASPC..261..551S}. Eclipse mapping of novalike systems such as UU~Aqr and IY~UMa has revealed enhanced emission along the ballistic stream beyond the initial impact site, consistent with such disk-skimming components \citep{2000MNRAS.314..713B}. In the helium-dominated ultracompact Gaia14aae, phase-resolve spectroscopy has revealed two bright spots, one consistent with the canonical stream--disk impact and a second likely associated with re-impact or vertically extended overflow \citep{2019MNRAS.485.1947G}. Comparable behavior has also been reported in other short-period systems such as SDSS~J0926$+$3624 \citep{2018MNRAS.478.3841S}. A modest contribution from analogous structures could contribute to the observed out-of-eclipse asymmetry in ATLAS~J1013$-$4516, although the present data do not uniquely require this interpretation. Alternative mechanisms---including magnetic channeling, disk warping, or anisotropic irradiation----could also produce phase-dependent asymmetries in disk emission. Given the degeneracy among these scenarios, the observed phase offsets primarily constrain the azimuthal location of enhanced emission rather than its detailed physical origin.

\subsection{Spectral Energy Distribution and Eclipse-Calibrated Decomposition} \label{sec:sed}

We constructed a broadband spectral energy distribution (SED) for ATLAS~J1013$-$4516 ultraviolet, optical, and near-infrared photometry from \textit{GALEX}, \textit{Swift}/UVOT, \textit{Gaia}, ULTRACAM, and VISTA, supplemented by an \textit{XMM-Newton} non-detection in the 0.2--2~keV band. The photometry spans $\sim$1500~\AA~ to $\sim$2.2~$\mu$m and represents the orbit-averaged system flux. Crucially, the ULTRACAM observations additionally provide wavelength-dependent measurements of the secondary-eclipse depth, which encode the relative contributions of the donor star and accretion flow as a function of wavelength. 

To exploit this wavelength dependence, we model the fractional eclipse depth in each band $i$ as
\begin{equation}
    \delta_i = f_{\rm occ}\frac{F_{{\rm donor},i} }{F_{{\rm donor},i}+F_{{\rm disk,i}}}
\end{equation}
where $F_{{\rm donor},i}$ and $F_{{\rm disk}, i}$ are the donor and disk fluxes synthesized through the corresponding bandpass, and $f_{\rm occ}$ is the fraction of the donor surface occulted during eclipse. This formulation directly links the eclipse morphology to the broadband SED and constrains the spectral shape of the eclipsed component. Assuming that the eclipsed light is dominated by the donor photosphere, the wavelength dependence of the eclipse depths tightly constrains the characteristic color temperature of the eclipsed component to a narrow range around $T_2=6550^{+70}_{-40}$~K, with uncertainties dominated by the choice of phase ranges used to estimate the eclipse depth due to the sensitivity to out-of-eclipse variability. This constraint arises from the relative eclipse depths across bands and is therefore insensitive to assumed system distance and plays a central role in decomposing the broadband SED.

We model the full SED using a single, self-consistent forward model that simultaneously reproduces the broadband fluxes and the secondary-eclipse depths. The model consists of two emitting components: a Roche-lobe--filling donor star and a luminous accretion disk, each approximated by a blackbody characterized by an effective temperature and emitting area. The donor radius is fixed by the donor mass under the assumption of Roche-lobe filling at the measured orbital period. The disk component is assigned an effective radius that absorbs uncertainties associated with its geometry, vertical structure, and radiative efficiency. For a given trial distance $d$, we obtain the line-of-sight reddening from the DECaPS three-dimensional dust map and allow for modest deviations to account for map uncertainties and small-scale structure. We adopt $R_V=3.32$ \citep{2016ApJ...821...78S} and apply a CCM-type extinction law in the ultraviolet and optical, while X-ray absorption is modeled via photoelectric attenuation.

The free parameters of the model are therefore
\[T_{\rm disk}, R_{\rm disk}, T_2, M_2, d, f_{\rm occ},\]
and the full parameter space is explored using nested sampling to capture parameter degeneracies and obtain posterior distributions.

Under the assumption that the eclipsed flux originates predominantly from the donor photosphere, the joint SED+eclipse fit yields a characteristic distance of $d=700^{+35}{-30}$~pc. This solution is internally self-consistent within the adopted forward model, combining the donor temperature constraint from the eclipse depths with the absolute flux normalization of the broadband SED. However, unlike the donor temperature, the inferred distance depends sensitively on assumptions about the spatial origin of the eclipsed light. Any additional red or optical emission that is spatially coincident with the donor during eclipse---such as reprocessed radiation from the irradiated donor face, vertically extended disk or stream--disk impact structures, or other azimuthally localized accretion-related emission---would bias the inferred donor flux upward and lead to a systematic underestimate the true distance. 

This sensitivity is highlighted by comparison to ES~Cet, which has a \emph{Gaia} parallax-constrained distance of $1.8\pm0.2$ kpc and a $G$-band magnitude of 16.8. 
The mass-transfer rate inferred for ATLASJ1013$-$4516 (Section~\ref{sec:evolution}) is comparable to that of ES~Cet, such that under the simplifying assumption of similar intrinsic accretion luminosities, the 2.7--mag difference apparent optical brightness would correspond to a characteristic distance of order $\sim6$ kpc for ATLAS~J1013$-$4516. This value is substantially larger than the distance inferred from our eclipse-calibrated SED, and the physical interpretation of this discrepancy is discussed in Section~\ref{sec:evolution}. 

\begin{figure}[ht!]
\center\includegraphics[width=0.47\textwidth]{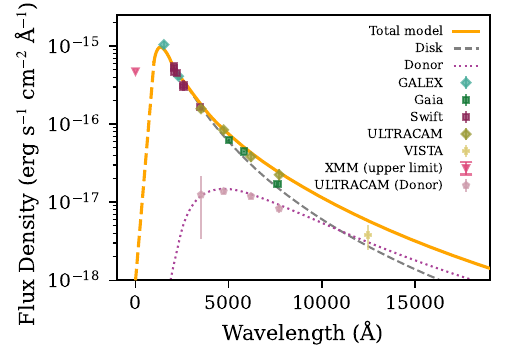}
    \caption{Spectral energy distribution of ATLAS~J1013$-$4516 with a representative two-component forward model drawn from the joint SED+eclipse posterior. The solid orange line shows the total model flux corrected for interstellar extinction, decomposed into contributions from the accretion disk (gray dashed line) and the Roche-lobe--filling donor star (purple dotted line). The pink points labelled ``ULTRACAM (Donor)" represent empirical donor flux estimates derived from the wavelength-dependent secondary-eclipse depths after correcting for the inferred occulted fraction of the donor light.
\label{fig:sed}}
\end{figure}

Therefore Figure~\ref{fig:sed} illustrates one such self-consistent SED solution under the assumption that the eclipsed flux originates is dominated by donor-like emission.

\subsection{Long-Term Timing Results} \label{sec:oc}

We jointly constrain the orbital ephemeris of ATLAS~J1013$-$4516 using photometry from \emph{Gaia}, ATLAS, ULTRACAM, and proto-Lightspeed, and report the resulting parameters in Table~\ref{tab:inferred}. Mid-eclipse times were measured from phase-folded subsets of the ATLAS and \emph{Gaia} light curves and combined with eclipse timings with the high-speed ground-based data.

\begin{figure}[ht!]
    \center\includegraphics[width=0.45\textwidth]{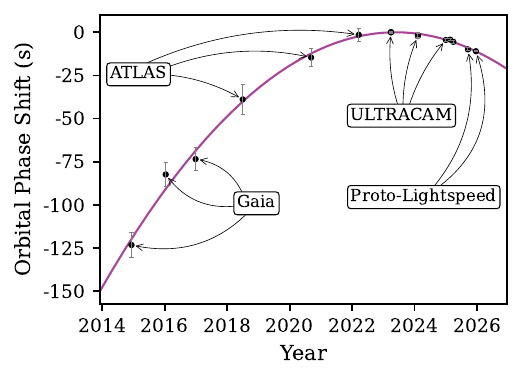}
    \caption{Observed minus calculated ($O-C$) diagram for ATLAS J1013$-$4516, showing mid-eclipse timing residuals from ATLAS, Gaia, ULTRACAM, and proto-Lightspeed photometry relative to a constant-period ephemeris.
    \label{fig:o-c}}
\end{figure}

To visualize deviations from a constant-period ephemeris, we construct an observed-minus-calculated ($O-C$) diagram, in which the difference between the observed mid-eclipse times and those predicted by a linear ephemeris is plotted as a function of observation year (Figure \ref{fig:o-c}). The timing residuals exhibit a clear, systematic drift relative to a constant-period model, motivating a quadratic ephemeris of the form
\begin{equation}
    T(E) = T_0 + P_0 E + \tfrac{1}{2}\dot{P}E^2
\end{equation}
where $T(E)$ is the predicted mid-eclipse time at integer cycle count $E$, $T_0$ is the reference epoch, $P_0$ is the orbital period at $T_0$, and $\dot{P}$ is the orbital-period derivative. 

Fitting this model yields
\[T_0=60033.2044878~\mathrm{BMJD_{TDB}}\pm0.327~\mathrm{s},\]
\[P_0=513.593303~\mathrm{s}\pm3.26~\mathrm{\mu s},\]
\[ \dot{P}=(-1.60\pm0.07)\times 10^{-12}~\mathrm{ s~s}^{-1}.\]
We adopt this refined ephemeris for all subsequent analysis.
\begin{table}
\raggedright
\begin{tabular}{l l} 
 \hline\hline
 $\alpha$ (ICRS, J2016.0) & 10:13:42.479 \\
 $\delta$ (ICRS, J2016.0) & $-45$:16:56.799 \\
 Parallax & $0.1992 \pm 0.2808$ mas \\
 $\mu_{\alpha*}$ & $-4.2974 \pm 0.2530$ mas yr$^{-1}$ \\
 $\mu_\delta$ & $+2.3348 \pm 0.2397$ mas yr$^{-1}$ \\
 Mean \emph{Gaia} $G$ & $19.5113 \pm 0.0104$ mag \\
 Epoch ($T_0$) & $60033.2044878~{\rm BMJD}_{\rm TDB}\pm 0.327~{\rm s}$\\
 Orbital period ($P$) & $513.593303\,{\rm s} \pm 3.26\,\text{\textmu}{\rm s}$ \\
 Period derivative ($\dot{P}$) & $(-1.60 \pm 0.07) \times 10^{-12}~\mathrm{s\,s^{-1}}$\\
 \hline\hline
\end{tabular}
\caption{Observed astrometric and orbital ephemeris parameters of ATLAS~J1013$-$4516.}
\label{tab:measured}
\end{table}

\section{Discussion} \label{sec:discussion}

In Table \ref{tab:inferred}, we summarize the parameters inferred from the measured period and its derivative, combined with physical constraints from arguments regarding gravitational wave-driven angular momentum loss, stable mass transfer, and the observed presence of an accretion disk.
\begin{table}
\centering
\begin{tabular}{l l}
\hline\hline
\multicolumn{2}{c}{\textbf{Evolution/dynamics constraints}}\\
\hline
Minimum accretor mass $M_{2\rm min}$ &
$0.52\,M_\odot$ \\

Minimum donor mass $M_{1, \rm min}$ &
$0.08\,M_\odot$ \\

Minimum donor radius $R_{2,\rm min}$ &
$0.027\,R_\odot$ \\

Accretor mass $M_1$ &
$0.87^{+0.36}_{-0.25}\,M_\odot$ \\

Donor mass $M_{2,\rm dyn}$ &
$0.10^{+0.03}_{-0.02}\,M_\odot$ \\


Distance $d$ (luminosity-scaling to ES~Cet) &
$\lesssim6\,\mathrm{kpc}$ \\

\hline
\multicolumn{2}{c}{\textbf{SED + eclipse constraints}}\\
\hline

Donor mass $M_{2, \rm SED}$  &
$0.160^{+0.018}_{-0.019}\,M_\odot$ \\

Donor temperature $T_2$ &
$6550^{+70}_{-40}\,\mathrm{K}$ \\

Accretion disk 
temperature $T_{\rm disk}$ &
$25{,}900^{+1{,}000}_{-1{,}100}\,\mathrm{K}$ \\

Distance, $d_{\rm SED}$ &
$\gtrsim 700\,\mathrm{pc}$ \\

\hline
\multicolumn{2}{c}{\textbf{Using combined constraints}}\\
\hline

4-yr \emph{LISA} SNR &
$12.3_{-5.2}^{+14.6}$ \\

Characteristic strain ($h_c$) &
$1.05_{-0.45}^{+1.24}\times10^{-20}$ \\




\hline\hline
\end{tabular}
\caption{Derived physical and evolutionary properties of ATLAS~J1013$-$4516. The dynamical donor mass is inferred from orbital-evolution constraints using $P$ and $\dot{P}$ under conservative mass transfer (Sec.~\ref{sec:pdotdiscussion}), while the SED-based donor mass and distance are derived from the joint SED+eclipse fit and depend on the assumptions about the origin of the eclipsed light (Sec.~\ref{sec:sed}). The distance inferred from an accretion luminosity scaling argument is described in Section \ref{sec:evolution}.}
\label{tab:inferred}
\end{table}

\subsection{Geometric and Dynamical Constraints} \label{sec:minmass}

The presence of an accretion disk in ATLAS~J1013$-$4516 imposes basic geometric constraints on the binary configuration. First, the donor must fill its Roche lobe in order to sustain mass transfer. At the observed orbital period of 8.56~min, this requires the donor's Roche-lobe radius to exceed the radius of a cold, fully degenerate white dwarf at the same mass. This condition sets a lower limit on the donor mass $M_2\gtrsim0.08\,M_\odot$, independent of assumptions about the mass-transfer rate. 

Second, the accretion flow must circularize to form a disk rather than directly impacting the accretor. For a given orbital period, this requires the white-dwarf accretor to be sufficiently compact that the circularization radius of the ballistic stream exceeds the accretor radius. This condition imposes a lower limit on the accretor mass, $M_1\gtrsim0.52\,M_\odot$, below which the system would instead operate in a direct-impact accretion regime. The methodology used to derive these constraints follows the geometric arguments presented by \citet{2024ApJ...977..262C}.

\begin{figure}[ht!]
\center\includegraphics[width=0.45\textwidth]{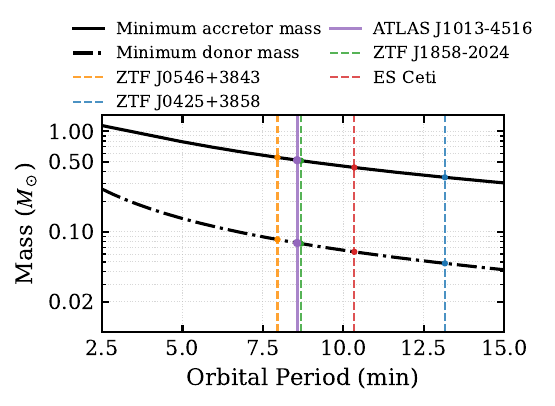}
\caption{Minimum component masses for ultracompact binaries as a function of orbital period, derived from geometric arguments that constrain the donor to fill its Roche lobe and the accretor to be compact enough to avoid direct-impact accretion. The curves follow the methodology outlined in \citet{2024ApJ...977..262C}. We overplot the constraints for the three systems discussed in that work (ZTF J0546+3843, ZTF J1858–2024, and ZTF J0425+3858), along with our new system, ATLAS J1013–4516, for which we find a minimum donor mass of $0.08\,M_\odot$ and minimum accretor mass of $0.52\,M_\odot$.
\label{fig:massconst}}
\end{figure}

Figure~\ref{fig:massconst} illustrates these constraints in the period-mass plane, compared with other ultracompact systems analyzed by \citet{2024ApJ...977..262C}. The resulting limits, summarized in Table~\ref{tab:inferred}, define a parameter space consistent with a massive white-dwarf accretor and a warm, semi-degenerate helium donor.

The X-ray non-detection provides an additional, independent constraint on the system geometry. Using the flux limits derived above and adopting the distance inferred from our optical modeling, the $0.2$–$12$~keV upper limit corresponds to an X-ray luminosity of $L_X \lesssim 6.8\times10^{30}$~erg~s$^{-1}$. For an accretor mass $M_1\simeq0.87\,M_\odot$ (allowing the range implied by our constraints), this corresponds to an Eddington-scaled luminosity of $L_X/L_{\rm Edd} \lesssim (4$–$9)\times10^{-8}$, well below the X-ray luminosities typically observed in short-period, disk-accreting AM~CVn systems with comparable inferred mass-transfer rates \citep{2005A&A...440..675R}. 

The absence of detectable X-ray emission therefore suggests that high-energy radiation from the inner accretion flow is strongly attenuated, most plausibly through geometric self-occultation in a high-inclination system. This interpretation is consistent with the deep optical eclipses and supports a picture in which the X-ray–emitting regions near the accretor are obscured by the disk rim or vertically extended accretion structures.

\subsection{Orbital Period Derivative and Donor Physics} \label{sec:pdotdiscussion}


Evaluated at the minimum component masses permitted by disk-mediated accretion ($M_1= 0.52\,M_\odot$, $M_2=0.08\,M_\odot$), gravitational radiation alone predicts
\begin{align}
\dot{P}_{\rm GR}
&= -\frac{96}{5}\,\frac{G^{5/3}}{c^5}
\left(\frac{2\pi}{P}\right)^{5/3}
\frac{M_1 M_2 (M_1+M_2)^{1/3}}{(M_1+M_2)^2} \notag\\
&\approx -5.51\times10^{-12}\ \mathrm{s\ s^{-1}} .
\label{eq:pdot_gr}
\end{align}
where we have adopted the standard formula for gravitational-wave-driven orbital decay of two point masses \citep{1964PhRv..136.1224P}. In contrast, the measured orbital-period derivative, $\dot{P}=(-1.60\pm0.07)\times 10^{-12}$~s~s$^{-1}$, is several times smaller in magnitude than the lower-limit expectation for a detached binary evolving under gravitational radiation alone.

In Roche-lobe--filling binaries, orbital angular momentum losses (e.g., gravitational radiation) act concurrently with mass transfer, which alters the component masses and thereby modifies the orbital response to a given angular-momentum loss. As the donor loses mass, its short-timescale structural response---quantified by the adiabatic mass-radius exponent $\xi_{\rm ad}\equiv \rm d\ln R_2/\rm d\ln M_2$---determines whether the donor expands or contracts, while the evolving mass ratio alters the Roche-lobe radius. Continued Roche-lobe overflow therefore couples angular momentum loss to mass transfer, such that for degenerate donors ($\xi_{\rm ad}<0$) the orbit can widen even as the total orbital angular momentum decreases.
This coupling gives rise to the characteristic period minimum in AM~CVn systems, separating an early phase of orbital decay from a later phase of orbital expansion once the donor's adiabatic response becomes sufficiently negative (e.g., \citealt{2001A&A...368..939N, 2003ApJ...598.1217D}).

The location of this period minimum---and more generally the net orbital-period derivative---additionally depends on the donor's long-term thermal and degeneracy evolution. We quantify this behavior using the evolutionary ratio $\chi_{\rm evol}$, defined as the ratio of the gravitational-radiation timescale to the donor's thermal timescale, such that systems with $\chi_{\rm evol}\ll 1$ have donors that rapidly radiate away thermal energy and approach a degenerate mass--radius relation on timescales short compared to the orbital inspiral, whereas systems with $\chi_{\rm evol}\sim 1$ retain significant thermal support throughout the gravitational-wave--driven evolution \cite{2024ApJ...977..262C}.
We therefore parameterize the orbital evolution using the set $(M_1, M_2, \xi_{\rm ad}, \chi_{\rm evol})$, adopting conservative mass transfer and assuming that angular-momentum losses are dominated by gravitational radiation. 

\begin{figure}[ht!]
\center\includegraphics[width=0.45\textwidth]{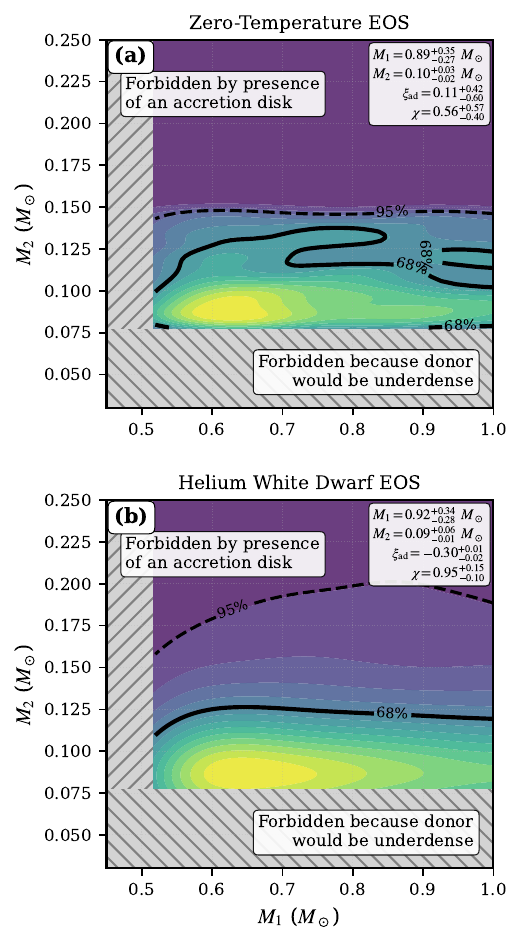}
\caption{
Comparison of mass and donor-physics constraints for ATLAS~J1013$-$4516 under different assumptions about the donor structure. 
Top: EOS-agnostic analysis in which the donor’s adiabatic mass–radius exponent $\xi_{\rm ad}$ and evolutionary ratio $\chi_{\rm evol}$ are treated as free parameters. 
Bottom: Finite-entropy helium white dwarf donor model based on the relations of \citet{2003ApJ...598.1217D}, which self-consistently link $(M_2, R_2, \xi_{\rm ad})$ and enforce Roche-lobe contact. 
Contours indicate 68\% and 95\% credible regions, while hatched areas denote excluded parameter space. 
While the inferred component masses are broadly consistent across both treatments, adopting a physically motivated donor equation of state substantially tightens the allowed ranges of $\xi_{\rm ad}$ and $\chi_{\rm evol}$, reflecting the reduced freedom in the donor’s mass–radius evolution.
\label{fig:pdotmassconst}
}
\end{figure}

Following this equation-of-state--agnostic parameterization, we perform a Bayesian analysis of $(M_1, M_2, \xi_{\rm ad}, \chi_{\rm evol})$ that simultaneously reproduces the observed $\dot{P}$ and geometric constraints (Figure~\ref{fig:pdotmassconst}). In this framework, the component masses are moderately well constrained, while the donor-physics parameters $\xi_{\rm ad}$ and $\chi_{\rm evol}$ remain weakly identified due to degeneracies between the donor's instantaneous radius response and its long-term thermal evolution. 

To assess the physical plausibility of the inferred parameter space, we repeat the analysis using finite-entropy helium white dwarf donor models, in which $(M_2, R_2, \xi_{\rm ad})$ are linked self-consistently through the finite-temperature relations of \citet{2003ApJ...598.1217D} and required to satisfy Roche-lobe contact. Under this assumption, the allowed ranges of $\xi_{\rm ad}$ and $\chi_{\rm evol}$ collapse to a comparatively narrow region of parameter space (Figure~\ref{fig:pdotmassconst}), indicating that if the donor is a warm helium white dwarf, the observed orbital-period derivative meaningfully constrains its effective radius response to mass loss. The inferred component masses remain broadly consistent with the EOS-agnostic and finite-entropy analyses, suggesting that the mass constraints are robust to uncertainties in the donor's thermal state; we therefore adopt the finite-entropy helium white dwarf model and summarize the corresponding marginalized posterior constraints in the Table~\ref{tab:inferred}.

Taken together, these results favor a semi-degenerate, thermally support helium donor that has not yet reached its minimum orbital period, while we emphasize that the tightening of the donor-physics parameters is conditional on the assumed equation of state. ATLAS~J1013$-$4516 therefore appears to occupy a transitional evolutionary regime consistent with systems approaching the turnover at which continued mass loss and donor expansion reverse the sign of $\dot{P}$.


\subsection{Population Context and Comparisons} \label{sec:evolution}

ATLAS J1013$-$4516 occupies a sparsely populated region of AM~CVn parameter space: disk-accreting ultracompact binaries with orbital periods shorter than 10~min. At such extreme orbital periods, it is remarkable that disk accretion can persist at all. The accretors in these systems are sufficiently compact that the circularization radius of the mass-transfer stream remains larger than the white-dwarf radius, allowing a disk to form even under intense gravitational-wave driving.

Even at the shortest orbital periods, the small population of ultracompact binaries exhibits striking diversity in their orbital evolution. Systems in this regime span both positive and negative orbital-period derivatives. ES~Ceti ($P\simeq10.3$~min; \citealt{2018ApJ...852...19D}) and ZTF~J1858$-$2024 ($P\simeq8.68$~min; \citealt{2024ApJ...977..262C}) show positive $\dot{P}$, consistent with systems that have passed their minimum period and are evolving toward longer periods. In contrast, ATLAS~J1013$-$4516 and ZTF~J0546$+$3843 \citep{2024ApJ...977..262C} exhibit a negative $\dot{P}$, indicating that they remain on the inspiral branch where gravitational-wave losses dominate over mass-transfer--driven expansion. Similarly negative $\dot{P}$ values are observed in HM Canri ($P\simeq5.4$~min; e.g., \citealt{2023MNRAS.518.5123M}) and V407 Vul ($P\simeq9.5$~min; \citealt{2004ApJ...610..416S}), both of which lack persistent accretion disks and operate in a direct-impact accretion regime. Taken together, these systems demonstrate that orbital evolution at the shortest periods is highly sensitive to donor structure, thermal state, and accretion geometry, even among binaries with comparable orbital periods.

To place ATLAS~J1013$-$4516 in an evolutionary context, we consider its mass-transfer rate under the minimum mass donor and accretor configuration. Using Equation~12 of \citealt{2024ApJ...977..262C}, we infer a mass-transfer rate of $\dot{M}\simeq6.8\times10^{-9}\,M_\odot\,\mathrm{yr}^{-1}$ for ATLAS~J1013$-$4516. This value is within an order of magnitude of estimates for ES~Cet, which has $\dot{M}=(2.5\pm1.6\times10^{-8}\,M_\odot\,\mathrm{yr}^{-1}$ and a \emph{Gaia} parallax--constrained distance of $1.8\pm0.2$~pc. At these orbital periods, the bolometric accretion luminosity is expected to scale approximately as $L_{\rm acc}\propto GM_1\dot{M}/R_1$, such that systems with comparable mass-transfer rates should have broadly similar intrinsic luminosities. Under this assumption, the $\sim$2.7-mag difference in apparent optical brightness between the two systems would correspond to a characteristic distance of order $\sim6$~kpc for ATLAS~J1013$-$4516, which is substantially larger than the distance inferred from eclipse-calibrated SED modeling. 

This apparent discrepancy may be reconciled if the observed optical flux of ATLAS~J1013$-$4516 is strongly suppressed by inclination-dependent obscuration or reprocessing. High-inclination accreting systems are known to exhibit reduced observed luminosities due to vertically extended disk rims, stream--disk impact regions, or disk winds that obscure or redistribute accretion power out of the optical bandpasses, an effect well documented in high-inclination X-ray binaries (e.g., \citealt{1982ApJ...257..318W}). Consistent with this interpretation, ES~Cet exhibits a partial eclipse of the outer accretion disk, indicative of a lower inclination, whereas the deep, sharp eclipses observed in ATLAS~J1013$-$4516 require a more edge-on geometry.
A higher inclination would naturally enhance self-obscuration, allowing a system with a high intrinsic accretion luminosity to appear optically faint.

Conversely, ultracompact systems viewed at low inclinations may lack strong orbital photometric modulation altogether, making them difficult to identify in variability-based searches. Together, these orientation-dependent effects introduce strong selection biases against both edge-on and face-on systems, potentially explaining the apparent scarcity of nearby ultracompact binaries even if such systems are intrinsically common.

In terms of light-curve morphology, ATLAS~J1013$-$4516 exhibits a pronounced post-eclipse hump whose amplitude increases toward bluer wavelengths. This feature occurs at a phase similar to that observed in other short-period disk accretors, including ES Ceti and ZTF~J1858$-$2024, and is naturally interpreted as emission from an azimuthally offset stream–disk impact region. Notably, the hump appears at an orbital phase earlier than is typically observed in hydrogen-accreting cataclysmic variables, where the hot-spot contribution often peaks immediately prior to primary eclipse (e.g., IP Pegasi; \citealt{2010MNRAS.402.1824C}). Among ultracompact AM~CVn systems themselves, the hump in ZTF J0546+3843 occurs at slightly earlier orbital phase than in ATLAS~J1013$-$4516, suggesting modest differences in disk size, mass-transfer rate, or stream trajectory even among systems with comparable orbital periods. 

What distinguishes ATLAS~J1013$-$4516 from other AM CVns at these orbital periods is its pronounced eclipsing geometry, including a sharp, kinked primary eclipse. This feature provides a precisely defined fiducial phase, enabling more precise long-baseline timing and a more tightly constrained orbital-period derivative. ATLAS~J1013$-$4516 also adds an important southern-hemisphere counterpart to the predominantly northern AM~CVn sample discovered by ZTF. Its detection in ATLAS highlights the value of all-sky cadence for identifying the shortest-period systems, where high-amplitude photometric variability remains accessible even to surveys with multi-night sampling.


\subsection{Gravitational-Wave Prospects} \label{sec:lisa}

With an orbital period of 8.56~min, ATLAS~J1013$-$4516 radiates gravitational waves at $f_{\rm GW}=2/P \simeq 3.9\,\mathrm{mHz}$, placing it well within the most sensitive band of \emph{LISA}. Adopting representative component masses within the range permitted by the eclipse geometry and orbital-evolution constraints (e.g., $M_1\simeq0.9\,M_\odot$, $M_2\simeq0.1\,M_\odot$) and a conservative uniform distance prior spanning $d\in [0.7,\,6]~\mathrm{kpc}$, we infer $h_c=1.05_{-0.45}^{+1.24}\times 10^{-20}$ and an expected four-year \emph{LISA} signal-to-noise ratio of $\mathrm{SNR}=12.3^{+14.6}_{-5.2}$, where the quoted uncertainties reflects the allowed distance range and mass uncertainties.

\begin{figure}[h!]
\center\includegraphics[width=0.45\textwidth]{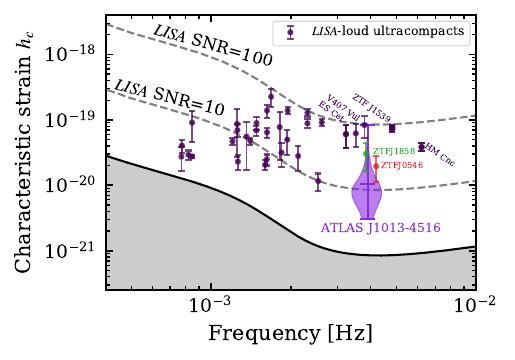}
\caption{The characteristic gravitational wave strain of ATLAS~J1013$-$4516 compared with the \emph{LISA} four-year sensitivity curve and selected \emph{LISA}-loud binaries \cite{2018MNRAS.480..302K,2020ApJ...905...32B,2024ApJ...977..262C}.
\label{fig:lisa}}
\end{figure}

Figure~\ref{fig:lisa} places ATLAS~J1013$-$4516 in the context of other known ultracompact binaries, showing that it lies comfortably above the four-year \emph{LISA} sensitivity curve and in a strain--frequency regime comparable to established verification binaries. While systems at similar frequencies span a wide range of evolutionary states, ATLAS~J1013$-$4516 stands out as a disk-accreting AM~CVn binary at sub-10~min periods with both eclipsing geometry and long-baseline timing constraints.

A \emph{LISA} detection would provide an independent measurement of the GW amplitude and orbital phase, as well as a constraint on the binary inclination. For an accreting system, the frequency derivative inferred from the GW signal does not directly yield the true chirp mass, since the observed orbital evolution is shaped by both gravitational-wave losses and mass transfer. Instead, comparison between GW-based constraints---particularly inclination---can be directly compared with the eclipse geometry and long-baseline electromagnetic timing to reduce degeneracies in interpreting the accretion-dominated light curve. In this way, ATLAS~J1013$-$4516 represents a particularly valuable target for joint GW--EM studies of ultracompact binary evolution.

\section{Conclusion} \label{sec:conclusion}

We have presented the discovery and multiwavelength characterization of ATLAS~J1013$-$4516, an 8.56~min disk-accreting AM~CVn binary with deep, multi-band eclipses and long-baseline orbital timing. The eclipsing geometry makes ATLAS~J1013$-$4516 an exceptional timing source, enabling precise measurements of orbital evolution in a regime where accretion-driven variability often limits timing fidelity in disk-accreting systems.

Using nearly a decade of ATLAS, \textit{Gaia}, ULTRACAM, and proto-Lightspeed observations, we measure a negative orbital-period derivative whose magnitude is reduced relative to the prediction for detached, gravitational-wave--driven inspiral. This indicates that mass ongoing mass transfer partially counteracts gravitational-wave angular-momentum losses, favoring a thermally inflated, semi-degenerate helium donor. Together with geometric constraints from eclipses, ATLAS~J1013$-$4516 provides a clean test of how donor structure influences secular orbital evolution at the shortest periods reached by disk-accreting white-dwarf binaries.

Recent discoveries have revealed a small but diverse population of ultracompact AM~CVn systems below $\sim$10 min, spanning a wide range of light-curve morphologies and both positive and negative orbital-period derivatives. ATLAS~J1013$-$4516 highlights the diversity while standing out as one of the few systems in this regime with deep eclipses and a precisely measured $\dot{P}$, making it a particularly valuable benchmark for testing models of disk-mediated mass transfer and helium-donor thermodynamics.

ATLAS~J1013$-$4516 is expected to be a detectable millihertz gravitational-wave source for \emph{LISA}. As one of the few known eclipsing \emph{LISA}-detectable binaries in the southern sky, it helps improve the sky distribution of well-characterized Galactic verification binaries. Continued eclipse timing will refine the orbital ephemeris and test for departures from a simple quadratic ephemeris, while future \emph{Gaia} data releases will extend homogeneous timing baselines that remain underutilized for ultracompact binary studies.

\begin{acknowledgments}
VSD and ULTRACAM are supported by STFC grant  ST/Z000033/1.
\end{acknowledgments}

%









\bibliography{ATLASJ1013}{}
\bibliographystyle{aasjournal}



\end{document}